\documentclass[a4paper,aps,pra,10pt,twocolumn,superscriptaddress,floatfix]{revtex4-1}

\usepackage{graphicx}
\usepackage{bm}
\usepackage{color}
\usepackage{xcolor,cancel}

\usepackage[colorlinks=true, pdfstartview=FitV, linkcolor=blue, citecolor=blue, urlcolor=blue]{hyperref}
\definecolor{shadecolor}{rgb}{0.95, 0.95, 0.86}

\begin{document}

\title{Universal Peregrine soliton structure in nonlinear pulse compression in optical fiber.}

\author{Alexey Tikan}
\affiliation{Laboratoire de Physique des Lasers, Atomes et Molecules,
  UMR-CNRS 8523,  Universit\'e de Lille, France}
\affiliation{Centre d'Etudes et de Recherches Lasers et Applications (CERLA), 59655
Villeneuve d'Ascq, France}

\author{Cyril Billet}
\affiliation{Institut FEMTO-ST,
CNRS Universit\'{e} Bourgogne-Franche-Comt\'{e} UMR 6174, 25030 Besan\c{c}on, France}

\author{Gennady El}
%\email{g.el@lboro.ac.uk}
\affiliation{Centre for Nonlinear Mathematics and Applications, Loughborough University, Department of Mathematical Sciences, Loughborough, LE11 3TU, UK}

\author{Alexander Tovbis}
%\email{Alexander.Tovbis@ucf.edu}
\affiliation{University of Central Florida, Department of Mathematics, USA}

\author{Marco Bertola}
% \email{marco.bertola@concordia.ca}
\affiliation{Department of Mathematics, Concordia University,  H3G 1M8, Montreal, Canada}
\affiliation{SISSA, Area of Mathematics, Trieste, Italy}

\author{Thibaut Sylvestre}
\affiliation{Institut FEMTO-ST,
CNRS Universit\'{e} Bourgogne-Franche-Comt\'{e} UMR 6174, 25030 Besan\c{c}on, France}

\author{Francois Gustave}
\author{Stephane Randoux}
\affiliation{Laboratoire de Physique des Lasers, Atomes et Molecules,
  UMR-CNRS 8523,  Universit\'e de Lille, France}
\affiliation{Centre d'Etudes et de Recherches Lasers et Applications (CERLA), 59655
Villeneuve d'Ascq, France}

\author{Go\"{e}ry Genty}
\affiliation{Tampere University of Technology, Department of Physics, Optics Laboratory, FI-33101 Tampere, Finland}

\author{Pierre Suret}
% \email{Pierre.Suret@univ-lille1.fr}
\affiliation{Laboratoire de Physique des Lasers, Atomes et Molecules,
  UMR-CNRS 8523,  Universit\'e de Lille, France}
\affiliation{Centre d'Etudes et de Recherches Lasers et Applications (CERLA), 59655
Villeneuve d'Ascq, France}

\author{John M. Dudley}
% \email{john.dudley@univ-fcomte.fr}
\affiliation{Institut FEMTO-ST,
CNRS Universit\'{e} Bourgogne-Franche-Comt\'{e} UMR 6174, 25030 Besan\c{c}on, France}

\date{\today}

\begin{abstract}
We present experimental evidence of the universal emergence of the Peregrine soliton predicted in the semi-classical (zero-dispersion) limit of  the focusing nonlinear Schr\"{o}dinger equation [Comm. Pure Appl. Math. {\bf 66}, 678 (2012)]. Experiments studying higher-order soliton propagation in optical fiber use an optical sampling oscilloscope and frequency-resolved optical gating to characterise intensity and phase around the first point of soliton compression and the results show that the properties of the compressed pulse and background pedestal can be interpreted in terms of the Peregrine soliton.  Experimental and numerical results reveal that the universal mechanism under study is highly robust and can be observed over a broad range of parameters, and experiments are in very good agreement with numerical simulations.

\end{abstract}

\maketitle

The focusing nonlinear Schr\"odinger equation (NLSE) is a fundamental model in nonlinear science that applies to a wide range of physical systems including water waves, plasmas, nonlinear fiber optics and Bose-Einstein condensates \cite{Akhmediev1997}. There is extensive current interest in the properties of particular NLSE solutions known as solitons on finite background, as they possess characteristics that suggest links with the emergence of rogue waves and extreme events on the ocean and in other environments \cite{Kharif2003,Onorato2013,Dudley2014}.  The prototype structure of this kind is the celebrated Peregrine soliton, which was first derived as a rational solution to the NLSE in the context of plane wave modulation instability \cite{Peregrine1983}.  Exciting the Peregrine soliton in experiments, however, requires careful choice of initial conditions, but a number of experiments have now observed its dynamics in the laboratory - first in nonlinear fiber optics \cite{Kibler2010}, and then in hydrodynamics \cite{Chabchoub2011} and  plasmas \cite{Bailung2011}.  These experiments have motivated much theoretical effort to study in detail the properties of the Peregrine soliton and other related NLSE solutions \cite{Akhmediev2013}.

It has been widely considered that the Peregrine
soliton was uniquely associated with the process of plane wave
modulation instability.  Recent theoretical studies,
however, have shown that it actually appears more generally as a
universal structure emerging during intensity localisation of high
power \textit{pulses} in the semi-classical (zero dispersion) limit
of the NLSE \cite{Bertola2012}.  Some evidence for this universality
has already been seen in experiments that
have characterised Peregrine soliton-like structures emerging during
partially coherent nonlinear wave propagation in optical fibers
\cite{Suret2016}.  But all other experimental studies of Peregrine
solitons have been restricted only to the regime of plane wave modulation
instability.

In a sense, this is surprising, because nonlinear propagation in optical fiber has been studied for decades \cite{Taylor1992,Agrawal2013}.  Moreover, high power pulse propagation in the anomalous dispersion regime of optical fiber is a well-known technique to generate pulse narrowing through higher-order soliton compression \cite{Mollenauer1983,Dianov1986,Agrawal2001}.  To our knowledge, however, any possible link between these compression dynamics and the localisation properties of the Peregrine soliton has never been explored.

In this paper, we fill this gap and present a detailed  theoretical
and experimental study that confirms the appearance of the universal
Peregrine soliton structure in nonlinear (N-soliton) pulse compression
in an NLSE system.  Our experimental results use two different
fiber optic based setups, and in one experiment where we
use the technique of frequency-resolved optical gating (FROG), we
fully characterize intensity and phase of the pulse during its
propagation. These results are not only of fundamental interest in showing the universality of the Peregrine soliton in cubic NLSE systems, but they also provide insight into the physical characteristics of nonlinear pulse compression.  In particular, nonlinear compression is well known to lead to a central temporally-localised peak sitting upon an extended pedestal \cite{Mollenauer1983}, and we are now able to interpret the physical origin of this structure in terms of the Peregrine soliton.

We begin by writing the focussing NLSE as follows:
\begin{equation}\label{NLS_semi}
i  \frac{\partial\psi}{\partial \xi} + \frac{1}{2N}\frac{\partial^2\psi}{\partial \tau^2}+ N |\psi|^2\psi = 0.
\end{equation}
\noindent Here, the envelope $\psi(\tau,\xi)$ is a function of
normalized distance $\xi$ and time $\tau$, and the input pulse is such
that $\psi(\tau,0)\to 0$ as $\tau \to \pm \infty$.  Note that these
initial conditions are fundamentally different to the plane wave used
to excite the Peregrine soliton from modulation instability
\cite{Dudley2009}. The  parameter $N>0$ (not  necessarily integer) that sets the dimensionless amplitude is well-known in optics as the soliton number ~\cite{Agrawal2013}, but the associated parameter $\epsilon=1/N$ is also used in descriptions of NLSE
dynamics, particularly in the semi-classical analysis ~\cite{Bertola2012}.

In terms of fiber parameters and physical
distance $z$ and time $t$, we have $\xi=z /
\sqrt{L_\mathrm{NL} \, L_\mathrm{D} }$ where we define nonlinear
length $L_\mathrm{NL} = (\gamma \mathrm{P}_0)^{-1}$, dispersive length
$L_\mathrm{D} = T_0^2/|\beta_2|$, and $\tau=t/T_0$.  Here $\gamma$ and
$\beta_2$ are the fiber nonlinearity and dispersion
respectively \cite{Agrawal2013}, and the input pulse is characterized
by timescale $T_0$ and peak power $P_0$.  The normalised envelope
$\psi = A/\sqrt{P_0}$ where $A(t,z)$ is the dimensional field (units
of $\mathrm{W}^{1/2}$).  The parameter $N  = \sqrt{L_\mathrm{D}/L_\mathrm{NL}} = \sqrt{\gamma P_0 T_0^2 /
  |\beta_2| }$ couples the fiber parameters and initial conditions.
When $N$ is an integer, the initial condition
$\psi(\tau,0)=\mathrm{sech}(\tau)$ represents an exact
  $N$-soliton solution of the NLSE, which follows the well-known periodic
  evolution~\cite{Yang}.

Analysis proceeds by writing the $\psi$ in terms of real variables (Madelung transform \cite{Madelung1926}) corresponding to intensity $\rho$ and instantaneous frequency (or chirp) $u$ defined through $\psi(\tau,\xi) = \sqrt{\rho(\tau,\xi)}\exp[-iN \int_{-\infty}^{\tau} u(\tau',\xi)d\tau']$.  Assuming a smooth (more precisely, analytic) initial pulse shape and large $N$, we  obtain a leading order approximation (nonlinear geometric optics) system:
\begin{equation}\label{NLS_displess}
\rho_\xi + (\rho u)_\tau = 0, \quad u_\xi + uu_\tau - \rho_\tau =0 \,
\end{equation}
describing the initial evolution of the pulse as long the derivative $\rho_\tau$ is not too large \cite{Kamvissis2003,Tovbis2004}.

It is known from analytical solutions of (\ref{NLS_displess})
  \cite{akhmanov_self_focusing_1966, gurevich_exact_1970}  and
  numerical  studies of the full NLSE (\ref{NLS_semi})
  \cite{Dudley2006} that nonlinear pulse propagation at high power
in the focussing NLSE typically leads to  temporal self-compression of
the intensity profile resulting, at some $\xi =
  \xi_\mathrm{c}$,  in a gradient catastrophe, the  point when the
  intensity profile has infinite derivative $|d\rho/d\tau| \to
  \infty$.  In the vicinity of this point the  geometric optics
  approximation (\ref{NLS_displess})  becomes invalid and the full
  NLSE must be used. Since $N$ is assumed to be large (equivalently,
  $\epsilon =1/N$ is small) one can take advantage of the semi-classical analysis of the NLSE inverse scattering solution performed in ref.~\cite{Bertola2012}. One of the key results of ref.~\cite{Bertola2012}  is that, when $\epsilon \ll 1$ (i.e. $N \gg 1$) the dynamics near the gradient catastrophe {\it universally} lead to the generation of the rational Peregrine soliton
 as a local asymptotic solution of the NLSE, which at the point of maximum compression $\xi_\mathrm{m}=\xi_c + O(N^{-4/5})$  assumes the form  $\psi_\mathrm{PS} (\tau,\xi_\mathrm{m}) = a_0 [ 1-4/(1 + 4 a_0^2 N^2 \tau^2) ] + O(N^{-1/5})$.

For the case of N-soliton initial data, the background $a_0$ for this local Peregrine soliton was shown in \cite{Tovbis2004} to be given by the field amplitude $\sqrt{\rho(0,\xi_c)}=\sqrt{2} + O(N^{-1/5})$ at the gradient catastrophe point found from solution to (\ref{NLS_displess}).  Significantly, the appearance of the Peregrine soliton  as a universal nonlinear coherent structure locally regularising the gradient catastrophe does not depend on  global properties (i.e. on the Zakharov-Shabat spectrum \cite{Zakharov-1972}) of the NLSE decaying initial condition  $\psi(\tau,0)$ -- it can have high soliton content or even be completely solitonless. Moreover, although this analysis is carried out in the semi-classical ($N \to \infty$) limit of the NLSE (which coincides with the nonlinear geometric optics approximation (2) only for $\xi < \xi_c$), as we shall see below from simulations and experiment, the appearance of the Peregrine soliton in the compressed pulse profile is extremely robust and is seen over a very wide range of $N$.

\begin{figure}[b!]
\centering
\includegraphics[width = 0.9 \linewidth]{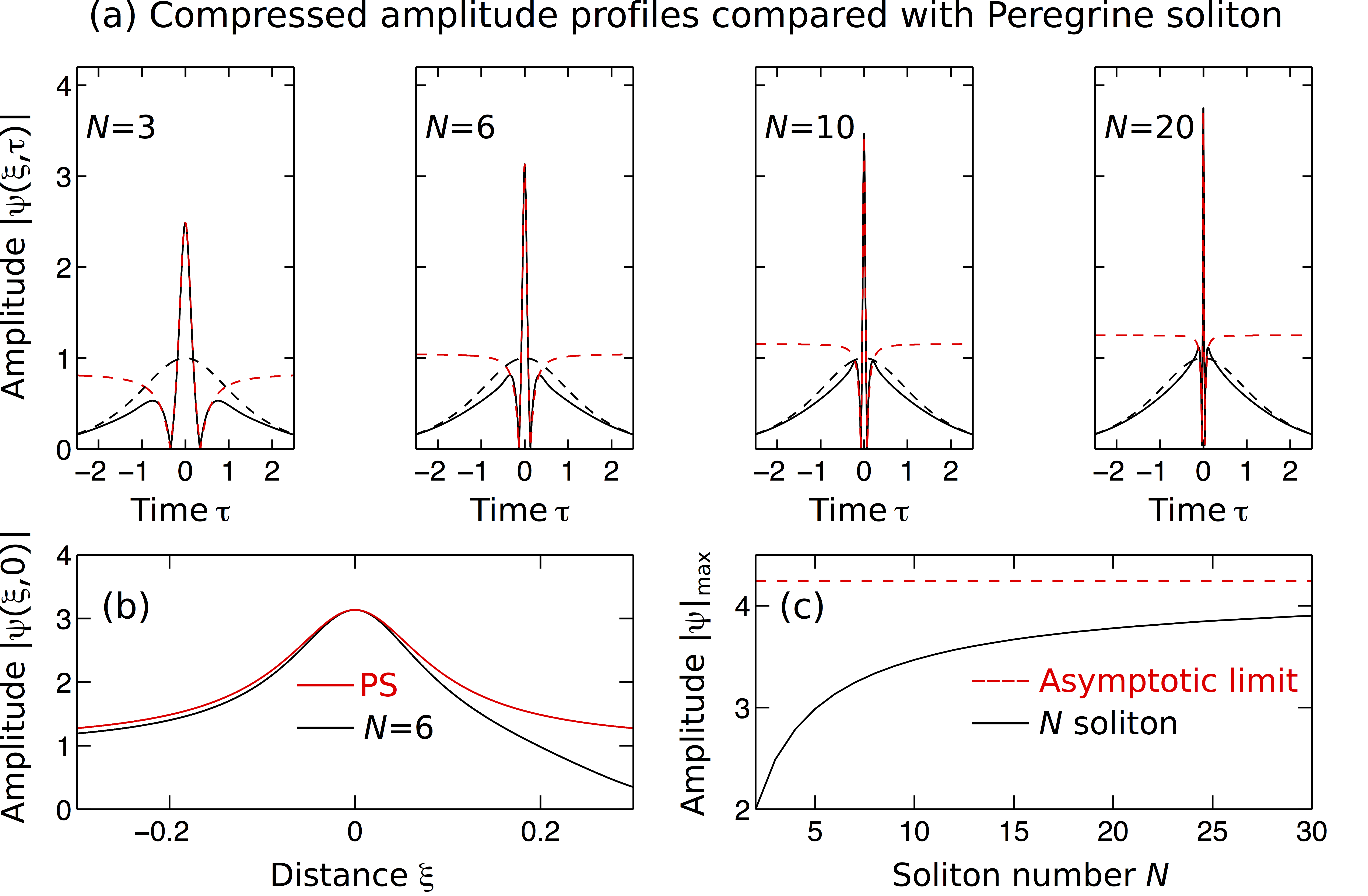}
\caption{(a) For different $N$ as shown, we compare pulse amplitude profiles at the compression point $\xi = \xi_\mathrm{m}$ (black solid line) with the intensity profile of a scaled Peregrine soliton (red dashed line). The black dashed line shows the input pulse $\psi(\tau,0) = \mathrm{sech}(\tau)$. (b) For $N=6$, we compare the longitudinal evolution of the amplitude of the evolving higher-order soliton and the Peregrine soliton at the temporal centre $\psi(\xi,0)$ (Note that the distance scale for the $N=6$ evolution is offset by $\xi_{m}$). (c) Numerical simulations showing the maximum amplitude $|\psi|_\mathrm{max}$  of the compressed higher-order soliton at the compression point as a function of $N$.  The asymptotic limit from semi-classical theory \cite{Bertola2012} ($3 \sqrt{2}\approx 4.24)$ is shown as the dashed line.}
\end{figure}

We begin by showing numerical results in Fig.~1 solving  Eq.~1 for pulsed initial condition $\psi(\tau,0) = \mathrm{sech}(\tau)$ with different $N$.  The nonlinear dynamics of high power pulse propagation in the focussing NLSE are well known, leading to a strongly-localised central peak surrounded by a temporally-extended pedestal at a compression distance  $\xi_\mathrm{m}$.  The solid black lines in Fig. 1~(a) show the amplitude at $\xi_\mathrm{m}$, with the black dashed line showing the profile of the input pulse.   The discussion above leads us to expect that the profile of the localised compressed peak will follow that of the Peregrine soliton, and indeed the analytic Peregrine soliton (dashed red line) is an excellent fit to simulation over the central region.

To further compare the properties of the compressed pulse and Peregrine soliton, Fig.~1(b)  compares the longitudinal evolution with $\xi$ of the amplitude $\psi(\xi,0)$ at the centre of the compressed pulse and the centre of the Peregrine soliton.  We note close agreement especially in the growth stage, but we remark that the decay stage of the pulsed evolution would be expected to deviate from the Peregrine soliton because of complex splitting dynamics beyond the compression point.  The maximum value of the pulse amplitude at the compression point described in \cite{Bertola2012} is denoted $|\psi_\mathrm{max}| = \psi(\xi_\mathrm{m},0)$, and simulation results plotting $|\psi_\mathrm{max}|$ as a function of $N$ are plotted in Fig.~1(c).  Significantly, in the large-$N$ limit, the amplitude approaches the asymptotic value of $ 3 \sqrt{2}$ as calculated in \cite{Bertola2012}.

\begin{figure}[!b]
\centering
\includegraphics[width = 0.9 \linewidth]{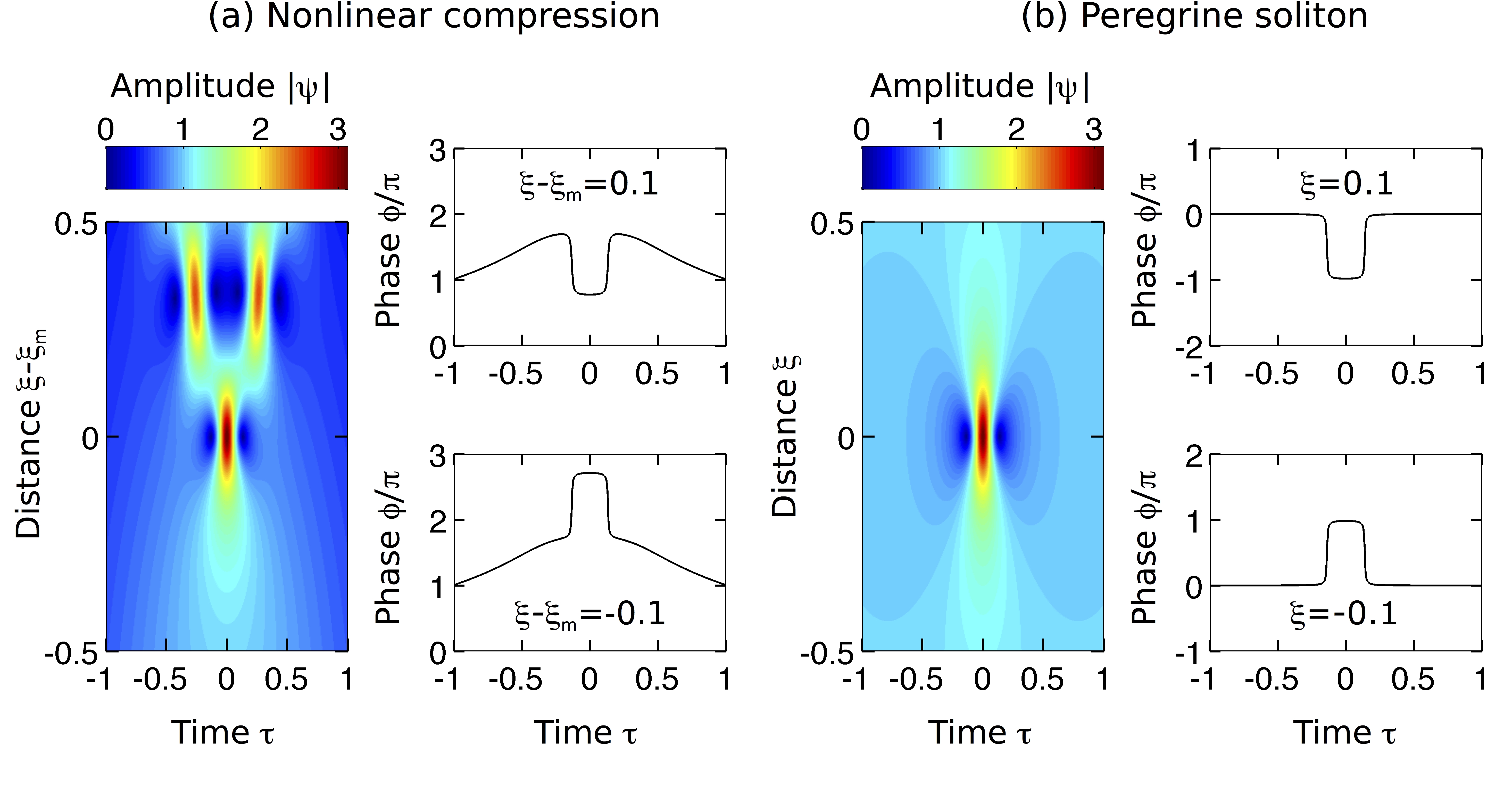}
\caption{Detailed view of amplitude and phase dynamics for (a) compression of a higher order $N=6$ soliton and (b) a Peregrine soliton. For clarity in the comparison, the distance scale in (a) is $\xi-\xi_m$ offset relative to the compression point.}
\end{figure}

To compare in more detail the compressed pulse characteristics with the Peregrine soliton, Fig. 2(a) shows an expanded view of the evolution from Fig.~1 in the vicinity of the compression point.  Fig. 2(b) shows the corresponding results for an ideal Peregrine soliton, and we see clearly the close similarity in the both amplitude and phase evolution. The confirmation of the expected  $\pi$ phase jump across the zero intensity points separating the ``wings'' and the central lobe of the Peregrine soliton is particularly striking.

These numerical studies show clearly how Peregrine soliton
characteristics appear {\it locally} during nonlinear
compression of a pulsed initial condition in an NLSE system.  We
stress again how fundamentally different this is from the emergence of
the Peregrine soliton in plane wave modulation instability.

We have confirmed these results experimentally
using two different setups injecting high power pulses in optical
fiber.  Fig. 3 shows a schematic of two setups used.  In setup 1, near-Gaussian pulses of $\Delta \tau =
5.3$~ps duration (FWHM) at 1525~nm from a spectrally-filtered optical
parametric oscillator (Coherent Chameleon) were injected into 400~m of
polarization maintaining fiber (Fibercore PMF).  The fiber parameters were $\beta_2 = -16.5 \times 10^{-27} \, \mathrm{s}^2 \, \mathrm{m}^{-1}$ and  $\gamma = 2.4 \times 10^{-3} \, \mathrm{W}^{-1} \, \mathrm{m}^{-1}$, and with injected pulse peak power of $P_0 = 3.3 \, \mathrm{W}$, we estimate $N \approx 2.2 $ for the input pulse ($\epsilon = 1/N \approx 0.45$). With the input pulse well-fitted by a Gaussian profile, the parameter $T_0 = \Delta \tau/1.665$ \cite{Agrawal2013}. Simulations were used to select the input power such that the fibre
length corresponded to close to the first compression distance, and
the pulse intensity profile  was measured using a custom-designed
optical sampling oscilloscope  (see  \cite{Walczak:15} and
Supplementary Information). The results obtained in this
case are displayed in Fig.~4  where we show the input pulse (green sampled points), the compressed pulse at the fibre output (blue sampled points), numerical simulation results (solid black line) and the ideal
theoretical Peregrine soliton (solid red line).  Note that there are
no free parameters used in the simulations.
%These results confirm the discussion above in showing how the central region of the compressed pulse intensity profile is very well-fitted by the Peregrine soliton profile.

\begin{figure}[!t]
\centering
\includegraphics[width = 0.9 \linewidth]{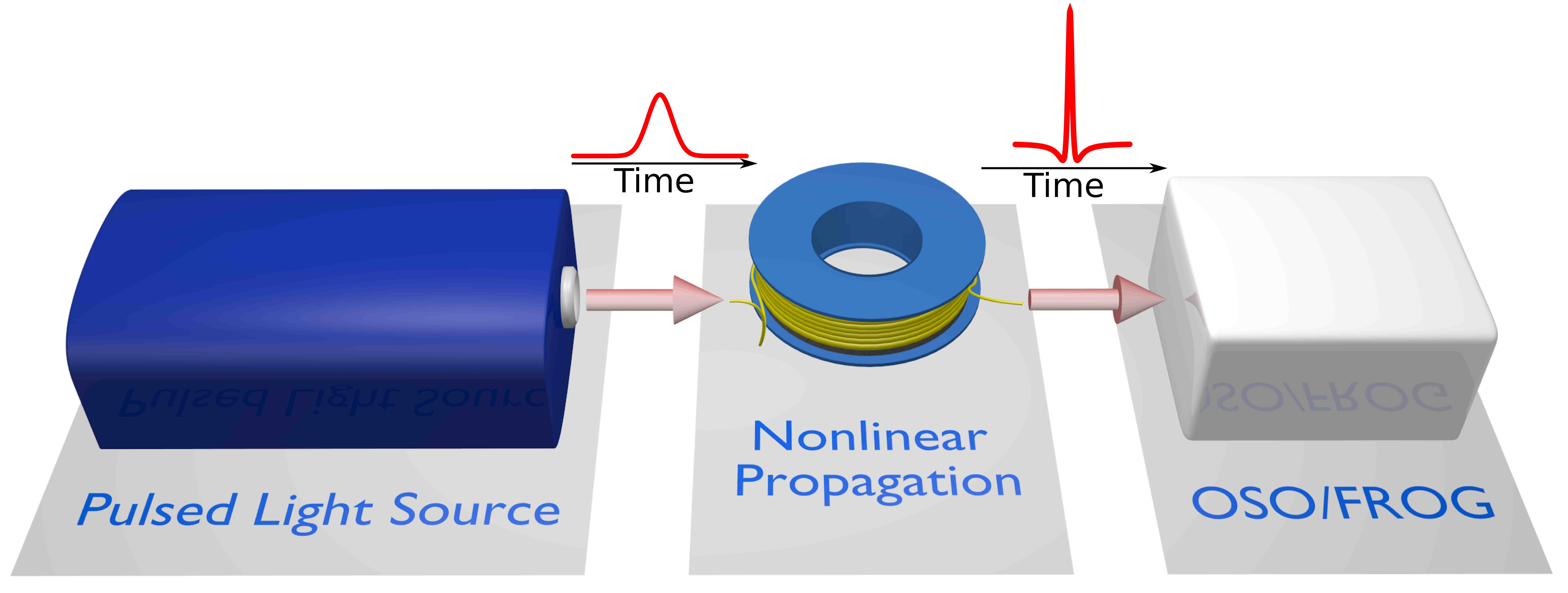}
\caption{Schematic Experimental setups. The pulsed light source is
  either a fiber picosecond laser either a spectrally filtered
  femtosecond OPO. The nonlinear propagation of pulses is achieved in
  a HNLF or in a standard PMF fiber}
\end{figure}

\begin{figure}[!h]
\centering
\includegraphics[width = 0.9 \linewidth]{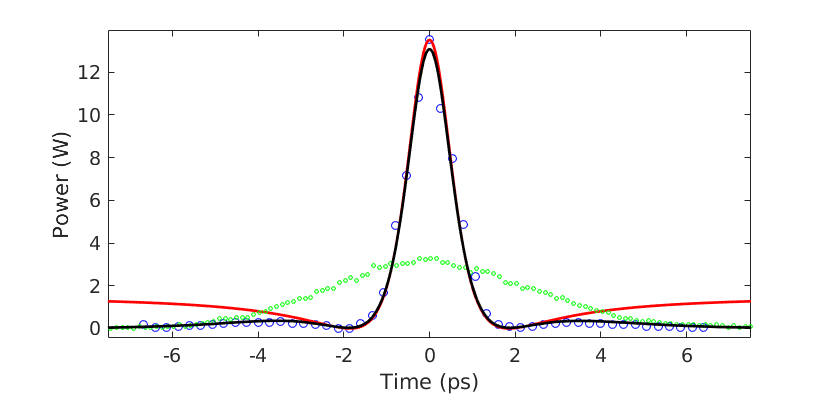}
\caption{Experimental and numerical simulations : temporal dynamics of
  the optical power (setup 1). Input pulse (green points)
  corresponding to $N=1/\epsilon\simeq2.2$. Output of
  the 400~m-long PMF (blue circles). Numerical simulations of NLSE (black
  line) and theoretical Peregrine soliton (red line). }
\end{figure}

\begin{figure*}[!t!]
\centering
\includegraphics[width = 0.9 \textwidth]{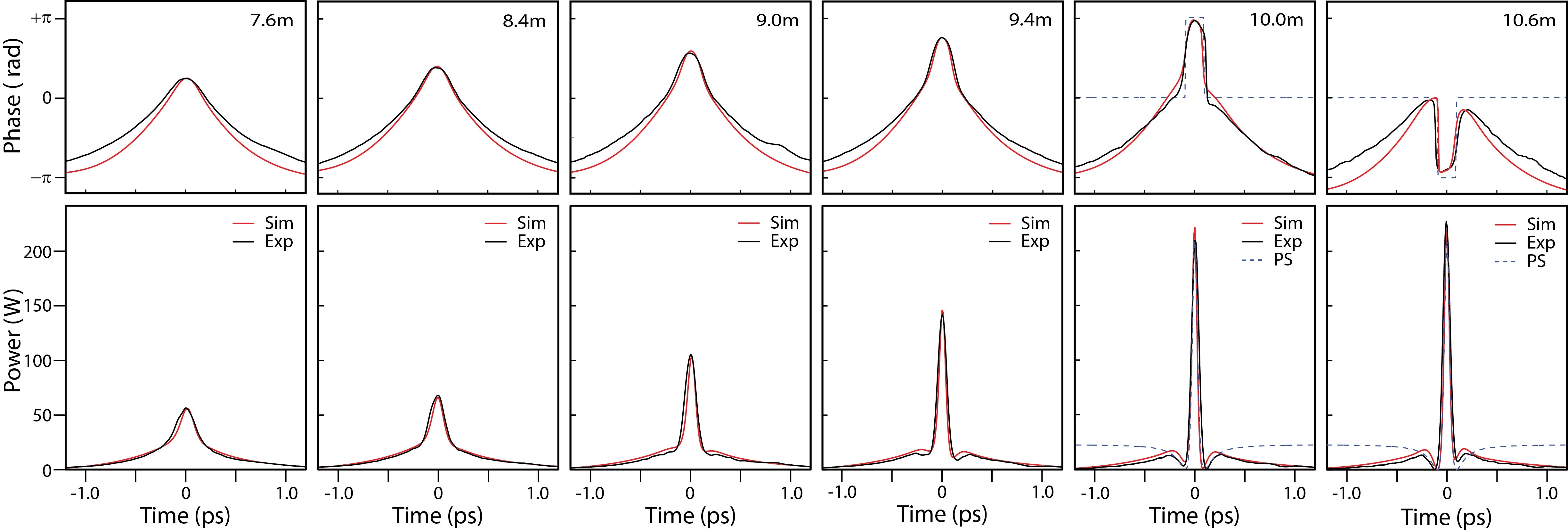}
\caption{Intensity (bottom) and phase (top) measurements of compressed pulse characteristics in optical fiber at the distances indicated, comparing experiment (black line) with simulations (red line).  For the results at 10.0~m and 10.6~m, we see clearly a flip in the phase characteristics across the central intensity lobe.  For these cases, we also plot the expected theoretical profile of an ideal Peregrine soliton.}
\end{figure*}

These experiments provide clear confirmation that nonlinear
pulse compression in optical fiber yields intensity characteristics in
good agreement with the Peregrine soliton.  To examine the pulse
properties in more detail, however, requires complete characterisation
of the compressed pulse in both amplitude and phase, and to this end
we developed a second experimental setup using second harmonic generation FROG to characterise the nonlinear pulse evolution \cite{Trebino2002,Dudley1999}.

In these experiments, the input pulses from a picosecond fibre laser (Pritel PPL) had duration $\Delta \tau =$~1.1~ps (FWHM) at 1550~nm.  The pulses were injected with input power $P_0 = 26.3$~W into Ge-doped highly nonlinear optical fiber (HNLF) fiber with $\beta_2 = -5.23 \times 10^{-27} \, \mathrm{s}^2\, \mathrm{m}^{-1}$ and $\gamma = 18.4 \times 10^{-3} \, \mathrm{W} \, \mathrm{m}^{-1}$.  The input pulses here were was well-fitted by a sech profile such that $T_0 = \Delta \tau/1.763$ \cite{Agrawal2013}, giving $N \approx 6$.  Numerical simulations for these parameters determined the first point of compression around $z = 10.3$~m.  Note that our simulations also included the effect of third order dispersion ($\beta_3 = 4.27 \times 10^{-41} \,\mathrm{s}^3\, \mathrm{m}^{-1}$), the Raman effect \cite{Agrawal2013} and input pulse asymmetry, but whilst including these effects was found to improve quantitative agreement with experiment (see below), the essential pulse dynamics up to the first compression point remain very well described by an ideal NLSE.  Of course, the higher-order effects play a major role beyond the compression point where they lead to supercontinuum broadening that prevents observation of recurrence to the initial state \cite{Dudley2006}.

Experiments were first performed at a maximum fiber length of 10.6~m, before the fiber was cut-back progressively.  At each fiber length, FROG measurements were made to yield intensity and phase profiles. FROG acquisition was performed on a $512 \times 512$ grid and FROG retrieval errors were typically $2 \times 10^{-3}$.  Standard checks of the retrieved pulse characteristics involving comparison with independent spectral and autocorrelation measurements were used to check measurement fidelity, and the direction-of-time ambiguity was lifted by an additional FROG measurement of propagation in a length of single mode fiber \cite{Trebino2002}.

The results of these experiments are shown in Fig.~5. Here we plot the retrieved intensity (bottom) and phase (top) at the fiber lengths indicated, comparing experiment (black line) with simulations (red line).  In all cases, we see excellent agreement between experiment and simulation, and we clearly observe the compression of the central region of the intensity profile and the development of a broader intensity pedestal.  The associated phase evolution is dominated by nonlinearity (where the phase profile follows the intensity profile) but as we approach the compression point, we see the development of a central region of phase with steepening edges upon a slower phase variation associated with the intensity pedestal.

Indeed for the results at 10~m and 10.6~m, we also plot the intensity and phase profile of an ideal Peregrine soliton solution, and there is excellent agreement with experiment and simulation across the pulse center.  Of course, one difference is that the Peregrine soliton background extends to $\tau \rightarrow \pm \infty$ whereas the pedestal observed in experiments is limited by the temporal width of the input pulse.  This highlights how the emergence of the Peregrine soliton here is a {\it local dynamical mechanism}. The qualitative  and quantitative agreement is clear, and in particular, as observed in numerical simulations of Fig. 2, the $\pi$ phase jump occurring at zero intensity  between the central lobe and the pedestal is a striking signature of the Peregrine soliton. Note finally, that the change of the sign of the phase derivative across the maximum compression point which is also a characteristic of  the analytic Peregrine soliton solution at the center of the pulse, is clearly observed in the experiments between 10~m and 10.6~m in Fig. 5.  Remarkably,  this property of the Peregrine soliton is also consistent with the generic change of sign of the phase derivative across the gradient catastrophe point \cite{gurevich_exact_1970}.

These results are very significant from both basic and applied viewpoints. From a fundamental perspective, our simulations and experiments confirm the predictions of Ref. \cite{Bertola2012} in showing how the regularization of the gradient catastrophe in the semiclassical NLSE leads to the emergence of the Peregrine soliton. Specifically, in addition to its appearance  in the development of plane wave modulational instability, we see how the Peregrine soliton also arises as the compressed pulse shape associated with high power nonlinear pulse propagation in the NLSE.   Our results reveal that this phenomenon arises over a very broad range of parameters : we locally observe the Peregrine soliton in the nonlinear regime of propagation as long as $\epsilon= \sqrt{L_\mathrm{D}/L_\mathrm{NL}}\leq 0.5 \, (N > 2)$.   For the particular system of nonlinear fiber optics, our results provide a clear physical interpretation to the long{standing observation that soliton-effect compression is necessarily associated with a broader pedestal, often considered deleterious for many applications.  Although techniques do exist which can reduce or suppress this pedestal to some degree \cite{Agrawal2001}, our results suggest that -- at least for nonlinear N-soliton compression scenarios described by the NLSE - its presence is inevitable.  Finally, we anticipate application of these results in studies of extreme events and rogue waves in integrable turbulence {\cite{Walczak:15,SotoCrespo:16,Suret2016,Randoux2016} where the problem of nonlinear localisation mechanisms remains a challenging open problem.

\begin{acknowledgments}
This work was supported by the European Research Council Advanced Grant ERC-2011-AdG-290562 MULTIWAVE and Proof of Concept Grant ERC-2013-PoC 632198-WAVEMEASUREMENT, the European Union Horizon 2020 research and innovation programme under grant agreement No 722380 SUPUVIR, the Agence Nationale de la Recherche (ANR OPTIROC ANR-12-BS04-0011, LABEX CEMPI ANR-11-LABX-0007 and ACTION ANR-11-LABX-0001-01), the Region of Franche-Comt\'{e} Project CORPS,  the French Ministry of Higher Education and Research, the Nord-Pas de Calais Regional Council and European Regional Development Fund (ERDF) through the Contrat de Projets Etat-R\'{e}gion (CPER Photonics for Society P4S). G. G. acknowledges the support from the Academy of Finland (Grants 267576 and 298463).
\end{acknowledgments}

% \bibliography{BibFile_parsed}

%

\end{document}